\input harvmac
\input epsf

\def\p{\partial}

\def\half{{1\over 2}}

\def\te{\tilde{\eta}}
\def\({\left(}
\def\){\right)}
%%%%%%%%%%%%%%%%%%%%%%%%%%%%%%%%%%%%%%%%%%%%%%%%%%%%%%%%%%%%%%%%%%%%%

\Title{}{\vbox{\centerline{Running Spectral Index in Noncommutative
Inflation}
\vskip7mm
\centerline{ and WMAP Three Year Results}}}

\centerline{Qing-Guo Huang and Miao Li } \bigskip
\centerline{\it
Institute of Theoretical Physics, Academia Sinica} \centerline{\it
P. O. Box 2735, Beijing 100080}
\medskip
\centerline{\it The interdisciplinary center of theoretical Studies,
Academia Sinica} \centerline{\it P. O. Box 2735, Beijing 100080}
\medskip
\centerline{\tt huangqg@itp.ac.cn} \centerline{\tt mli@itp.ac.cn}

\bigskip
A model independent analysis shows that the running of the
spectral index of the three year WMAP results can be nicely
realized in noncommutative inflation. We also re-examine some
concrete noncommutative inflation models. We find that a large
tensor-scalar ratio is required, corresponding to a low number of e-folds
before the end of inflation in some simple models.

\Date{March, 2006}
%\draft

\nref\infl{A.H. Guth, Phys.Rev.D 23(1981)347; A.D. Linde,
Phys.Lett.B 108(1982)389; A. Albrecht and P.J. Steinhardt,
Phys.Rev.Lett. 48(1982)1220. }

\nref\wmapts{D.N. Spergel et. al, astro-ph/0603449.}

\nref\al{L. Alabidi and D.H. Lyth, astro-ph/0603539. }

\nref\cst{D.J.H. Chung, G. Shiu, M. Trodden, Phys.Rev.D
68(2003)063501, astro-ph/0305193; S.M. Leach and A. Liddle,
Phys.Rev.D 68(2003)123508, astro-ph/0306305.}

\nref\ncst{T. Yoneya, in " Wandering in the Fields ", eds. K.
Kawarabayashi, A. Ukawa ( World Scientific, 1987), p. 419; M. Li and
T. Yoneya, Phys. Rev. Lett. 78 (1977) 1219; T.
Yoneya, Prog. Theor. Phys. 103, 1081 (2000), hep-th/0004074; J.
Polchinski, " String Theory " volume 2; J. She, hep-th/0509067,
hep-th/0512299.}

\nref\bh{R. Brandenberger, P.M. Ho, Phys.Rev.D 66(2002) 023517,
hep-th/0203119. }

\nref\hlf{Q.G. Huang and M. Li, JHEP 0306(2003)014, hep-th/0304203.
}

\nref\hle{Q.G. Huang and M. Li, JCAP 0311(2003)001,
astro-ph/0308458. }

\nref\hlp{Q.G. Huang and M. Li,  Nucl.Phys.B 713(2005)219,
astro-ph/0311378. }

\nref\pncf{S. Tsujikawa, R. Maartens and R. Brandenberger,
astro-ph/0308169; E. Grezia, G. Esposito, A. Funel, G. Mangano, G.
Miele, gr-qc/0305050; M. Fukuma, Y. Kono, A. Miwa, hep-th/0307029;
D. Liu, X. Li, astro-ph/0402063; H. Kim, G. Lee, Y. Myung,
hep-th/0402018; H. Kim, G. Lee, H. Lee, Y. Myung, hep-th/0402198; R.
Cai, hep-th/0403134; G. Calcagni, hep-th/0406006; S. Alavi, F.
Nasseri, astro-ph/0406477; Y. Myung, hep-th/0407066; G. Barbosa, N.
Pinto-Neto, hep-th/0407111; G. Calcagni, S. Tsujikawa,
astro-ph/0407543; G. Barbosa, hep-th/0408071; D. Liu, X. Li,
hep-th/0409075; K. Bamba, J. Yokoyama, hep-ph/0409237; K. Bamba, J.
Yokoyama, hep-ph/0502244; G. Calcagni, hep-ph/0503044; C-S. Chu, B.
R. Greene and G. Shiu, hep-th/0011241; F. Lizzi, G. Mangano, G.
Miele and M. Peloso, JHEP 0206 (2002) 049, hep-th/0203099. }

\nref\prunt{K. Izawa, hep-ph/0305286; S. Cremonini, hep-th/0305244;
E. Keski-Vakkuri, M. Sloth, hep-th/0306070; M. Yamaguchi, J.
Yokoyama, hep-ph/0307373; J. Martin, C. Ringeval, astro-ph/0310382;
K. Ke, hep-th/0312013; S. Tsujikawa, A. Liddle, astro-ph/0312162; C.
Chen, B. Feng, X. Wang, Z. Yang, astro-ph/0404419; L. Sriramkumar,
T. Padmanabhan, gr-qc/0408034; N. Kogo, M. Sasaki, J. Yokoyama,
astro-ph/0409052; G. Calcagni, A. Liddle, E. Ramirez,
astro-ph/0506558; H. Kim, J. Yee, C. Rim, gr-qc/0506122. }

\nref\pruns{A. Ashoorioon, J. Hovdebo, R. Mann, gr-qc/0504135; G.
Ballesteros, J. Casas, J. Espinosa, hep-ph/0601134; J. Cline, L.
Hoi, astro-ph/0603403. }

\nref\run{B. Feng, M. Li, R. J. Zhang and X. Zhang,
astro-ph/0302479; J. E. Lidsey and R. Tavakol, astro-ph/0304113; L.
Pogosian, S. -H. Henry Tye, I. Wasserman and M. Wyman, Phys. Rev. D
68 (2003) 023506, hep-th/0304188; S. A. Pavluchenko,
astro-ph/0309834; G. Dvali and S. Kachru, hep-ph/0310244. }

\nref\alsl{A.R. Liddle, S.M. Leach, Phys.Rev.D 68(2003)103503. }

An epoch of accelerated expansion in the early universe, inflation,
dynamically resolves many cosmological puzzles such as homogeneity,
isotropy and flatness of the universe \infl, and generates
superhorizon fluctuations without appealing to fine-tuned initial
setups. These fluctuations become classical after crossing out the
Hubble horizon. During the deceleration phase after inflation they
re-enter the horizon, and seed the matter and the radiation
fluctuations observed in the universe. The anisotropy in CMB encodes
the very important information for inflation.

The $\Lambda$CDM model remains an excellent fit to the three years
WMAP data  and other astronomical data \wmapts. In the simplest
models for the structure formation, a scale-invariant spectrum of
the primordial power spectrum is no longer a good fit to the three
year data, implying that inflation must be dynamic. Some simple
inflation models are already ruled out \refs{\wmapts, \al}. The
deficit of power in  low mulitipoles of the angular power spectrum
in the first year results of WMAP is not preferred by the three
years results anymore. Even though allowing for a running spectral
index slightly improves the fit to the WMAP data, the improvement in
the fit is not significant enough to require this new parameter.
However, it is important that a running index still survives  the
three year data together with other sets of data. Since it is hard
to realize a large running in the usual slow-roll inflation models
\cst, we hope that future experiments will confirm a running
spectral index and thus open a new window into the physics of the
first moments of the big bang and provides some  clues into
trans-Planckian physics.

Noncommutative spacetime naturally emerges in string theory \ncst,
which implies a new uncertainty relation \eqn\urst{\Delta t_p \Delta
x_p \geq l_s^2, } where $t_p$ and $x_p$ are the physical time and
space, $l_s$ is uncertainty length scale or string scale in string
theory. Constructing a realistic inflation model in noncommutative
spacetime is still an open question. Motivated by the first year
WMAP resuls, we found in \refs{\hlf-\hlp} that a toy noncommutative
inflation model \bh\ can accommodate  a large running, this model
was later extensively studied in \pncf. Particularly in \hlp\ we
found that the noncommutative effects always make the power spectrum
more blue and the noncommutative effects on the small scale
fluctuations can be ignored, which is consistent with observational
results. Other models with a large running are discussed in
\refs{\prunt-\run}. We re-examine the model studied in
\refs{\hlf-\hlp} in the light of the three year WMAP results.

The spacetime noncommutative effects are encoded in a new product among
functions,
namely the star product, replacing the usual algebra product.
The evolution of the background is homogeneous and the standard
cosmological equations of the inflation will not change and still
take the form in Friedmann-Robertson-Walker (FRW) Universe: \eqn\mp{
\ddot \phi + 3 H \dot \phi + V'(\phi) = 0, } \eqn\hu{ H^2 =
\left({\dot a \over a} \right)^{2} = {1 \over 3 M_p^2} \left( \half
{\dot \phi}^2 + V(\phi) \right), } here $M_p$ is the reduced Planck
mass and we assumed the universe be spatially flat and the inflaton
$\phi$ be spatially homogeneous. If $\dot \phi^2 \ll V(\phi)$ and
$\ddot \phi \ll 3 H \dot \phi$, the scalar field slowly rolls down
its potential. Define some slow-roll parameters, \eqn\ep{\epsilon_V=
{M_p^2 \over 2} \left({V' \over V}\right)^2, } \eqn\et{\eta_V =M_p^2
{V'' \over V}, } \eqn\xx{\xi_V=M_p^4 {V' V''' \over V^2}. } The
slow-roll condition can be expressed as $\epsilon_V, \eta_V \ll 1$.

In order to make the uncertainty relationship in \urst\ more clear
in the Friedmann-Robertson-Walker (FRW) background, we introduce
another time coordinate $\tau$ in the noncommutative spacetime such
that the metric takes the form
\eqn\mtt{ds^2=dt^2-a^2(t)d{\vec{x}}^2=a^{-2}(\tau)d\tau^2-
a^2(\tau)d{\vec{\tau}}^2.} Now the uncertainty relationship \urst\
becomes \eqn\ustc{\Delta \tau \Delta x \geq l_s^2. } The star
product can be explicitly defined as
\eqn\dstp{f(\tau,x)*g(\tau,x)=e^{-{i\over 2}l_s^2(\p_x \p_{\tau'} -
\p_{\tau} \p_{x'})}f(\tau, x)g(\tau', x')|_{\tau'=\tau, x'=x}.}
Since the comoving curvature perturbation $\cal R$ depends on the
space and time, the equation of motion for $\cal R$ is modified by
the noncommutative effects \eqn\emsp{u_k''+\(k^2-{z_k'' \over z_k}
\)u_k=0, } where \eqn\scaf{\eqalign{z_k^2(\te)&=z^2y_k^2(\te), \quad
y_k^2=(\beta_k^+\beta_k^-)^{\half},\cr {d\te \over
d\tau}&=\left({\beta_k^-\over \beta_k^+}\right)^\half,\quad
\beta_k^\pm =\half (a^{\pm 2}(\tau+\l_s^2k)+a^{\pm
2}(\tau-l_s^2k)),}} here $l_s$ is the string length scale,  $z = a
\dot \phi / H$, ${\cal R}_k(\te) = u_k(\te) / z_k(\te)$ is the
Fourier modes of $\cal R$ in momentum space and the prime denotes
derivative with respect to the modified conformal time $\te$. The
deviation from the commutative case encodes in $\beta^\pm_k$ and the
corrections from the noncommutative effects can be parameterized by
${Hk\over aM_s^2}$. After a lengthy but straightforward calculation,
we get \eqn\zkt{\eqalign{{z_k'' \over z_k}&=2(aH)^2\(1+{5\over
2}\epsilon_V-{3\over 2}\eta_V-2\mu \), \cr aH&\simeq {-1\over
\te}(1+\epsilon_V+\mu),}} where $\mu = {H^2 k^2 / (a^2 M^4_s)}$ is
the noncommutative parameter and $M_s = l_s^{-1}$ is the string mass
scale. Solving eq. \emsp\ yields the amplitude of the scalar
comoving curvature fluctuations in noncommutative spacetime
\eqn\asc{\Delta_{\cal R}^2\simeq {k^3\over 2\pi^2}\left|{\cal
R}_k(\te) \right|^2={V/M_p^4 \over 24\pi^2\epsilon_V}
(1+\mu)^{-4-6\epsilon_V+2\eta_V}, } where $H$ and $V$ take the
values when the fluctuation mode $k$ crosses the Hubble radius
($z_k''/z_k=k^2$), $k$ is the comoving Fourier mode. Plugging this
condition into eq. \zkt, we obtain \eqn\dku{\eqalign{d\ln
k&=(1-\epsilon_V+4\epsilon_V \mu)Hdt, \cr {d\mu \over d\ln
k}&=(1+\epsilon_V-4\epsilon_V\mu){1 \over H}{d\over dt}\({H^2k^2
\over a^2M_s^4} \)\simeq -4\epsilon_V\mu.}} Using eq. \asc\ and
\dku, we obtain the index of the power spectrum for the scalar
fluctuations and its running respectively \eqn\pe{n_s-1 \equiv s={d
\ln \Delta_{\cal R}^2 \over d\ln k }
=-6\epsilon_V+2\eta_V+16\epsilon_V\mu, } \eqn\er{{d n_s \over d\ln
k}\equiv \alpha_s=-24\epsilon_V^2+16\epsilon_V \eta_V-2\xi_V
-32\epsilon_V\eta_V\mu. } The tensor-scalar ratio is
\eqn\tsr{r=16\epsilon_V. } Here we only sketch out the brief
deviation of the primordial power spectrum and the spectral index
and its running. See \hlp\ in details. With WMAP data only, the best
fit results for the $\Lambda$CDM model with running and tensors are
\eqn\wm{\eqalign{n_s=1.21^{+0.13}_{-0.16}, \quad dn_s/d\ln
k=-0.102^{+0.050}_{-0.043}, \cr \Delta_{\cal R}^2(k=0.05
Mpc^{-1})=\(20.9_{-1.9}^{+1.3} \)\times10^{-10},}} and
\eqn\scw{r\leq 1.5 \quad \hbox{at 95$\%$ CL}.} We need to stress
that the improvement in the fit is not significant enough to require
a new parameter, even through a running spectral index slightly
improves the fit to the WMAP data. Since $\epsilon_V$ must be
positive, eq. \pe\ says that the noncommutative effects always make
the spectral index more blue. If $\eta_V$ is also positive, the
noncommutative effects can help us to fit the spectral index and its
running nicely.

From eq.\pe, we have $\eta_V=\half\(s+{3\over 8}r-r\mu \)$. With
this substituted into eq.\er, it becomes
\eqn\ers{{d n_s \over d \ln
k}={3 \over 32}r^2+{s \over 2}r-r\({7\over
8}r+s\)\mu+r^2\mu^2-2\xi_V. } If spacetime is commutative ($\mu =
0$), eq.\ers\ becomes \eqn\aer{{d n_s \over d \ln k}={3 \over
32}r^2+{s \over 2}r-2\xi_V. }
Thus in the commutative case, $\xi_V$ must
be large in order to get a large enough negative value of $d
n_s / d \ln k$ in \aer, when the CMB power spectrum is
blue $(s
> 0)$,  and $r>0$. However it is quite hard
to have a large $\xi_V$ in the known typical inflation model \cst. If
we take the noncommutative effects into account, the second or the
third term on the right hand side of eq.\ers\ will become negative
for some suitable values of $\mu$, there is no longer the need for a
large $\xi_V$.

Ignoring $\xi_V$, the running $\alpha_s$ becomes negative provided
\eqn\ineq{r>{\mu s-{s\over 2}\over {3\over 32}-{7\over 8}\mu +
\mu^2},} where we assumed that the denominator is negative, this  is
the case when $r$ is in the range ${1\over 8} <\mu <{3\over 4}$. If
$\mu>{1\over 2}$, and $s$ is positive (which is the case at $k=0.002
(\hbox{Mpc})^{-1}$), the lower bound for $r$ is negative thus in
this case the running is always negative. For a more reasonable
value of $\mu$, $\mu<{1\over 2}$,  the lower bound for $r$ is
positive. As an example, taking $s=0.2$, $\mu ={1\over 4}$, the
lower bound for $r$ is $0.8$. We see that indeed for a small $\mu$,
a large $r$ is required to have a negative running.

Since the value of $\xi_V$ depends on the concrete inflation model,
we shall ignore $\xi_V$ which is positive in many typical inflation
models in order to make a model-independent analysis and show the
constraint on values of $r=16\epsilon_V$ and of $\mu$ in order to
fit the experimental data, in fig. 1.

\bigskip
{\vbox{{\epsfxsize=9cm
        \nobreak
    \centerline{\epsfbox{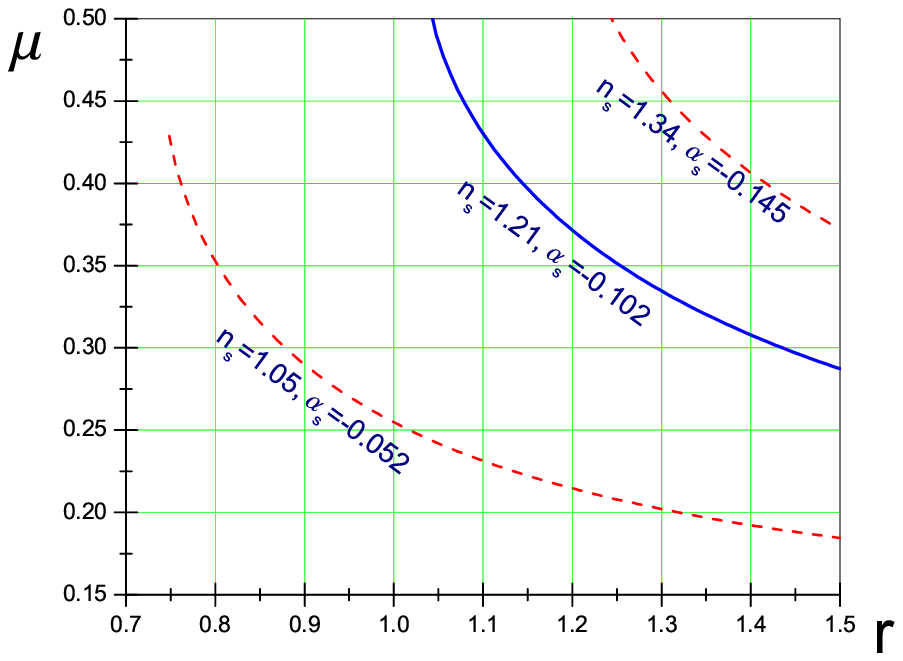}}
        \nobreak\bigskip
    {\raggedright\it \vbox{
{\bf Figure 1.} {\it The value for the blue solid line is $n_s=1.21,
{d n_s / d \ln k}=-0.102$. The tensor-scalar ratio $r$ is not
greater than 1.5 at $95\%$ CL. The range between these two red dash
lines is permitted. Here we neglect $\xi_V$.}
 }}}}
    \bigskip}

Large enough tensor-scalar ratio is needed $(r\geq 0.75)$ in order
to fit the running of the spectral index for WMAP. The permitted
range for the parameter is nice and our results can be trusted.

\medskip
\noindent $\bullet$ {\it Chaotic inflation}

Chaotic inflation is drived by the scalar field with potential
$V(\phi)\sim \phi^n$. The slow roll parameters are related to the
number of e-folds $N$ before the end of the inflation by
\eqn\cisp{\eqalign{\epsilon_V&={n\over 4N}, \cr \eta_V&={n-1 \over
2N}, \cr \xi_V &={(n-1)(n-2)\over 4N^2}.}} The spectral index and
its running and the tensor-scalar ratio are given by
\eqn\cir{\eqalign{n_s&=1+\(\mu-{1\over 8}-{1\over 4n} \)r, \cr
\alpha_s&=-{n+2\over 32n^2}r^2\(1+{8n(n-1)\over n+2}\mu \), \cr
r&={4n\over N}. }} The running of the spectral index is always
negative for the chaotic inflation model. The power spectrum is
always a red spectrum in the commutative spacetime. But it becomes
blue if $\mu \geq {1\over 8}+{1 \over 4n}$ in noncommutative chaotic
inflation. In the following we check two special cases with $n=2$
and $n=4$.

\noindent 1) For n=2.

The spectral index and its running and the tensor-scalar ratio are
\eqn\cirt{\eqalign{n_s&=1+\(\mu-{1\over 4}\)r, \cr \alpha_s&=-{r^2
\over 32}\(1+4\mu \), \cr r&={8\over N}.}} If we take $14\leq N \leq
75$ \refs{\al, \alsl}, $0.11\leq r \leq 0.57$ and $-1.0\times
(1+4\mu)\times 10^{-2}\leq \alpha_s \leq -3.8\times (1+4\mu)\times
10^{-4}$. In the commutative spacetime $\mu=0$, the running of the
spectral index satisfies $\alpha_s\in (-10^{-2}, -10^{-4})$. If we
take $N=14$ to fit $n_s=1.21$, we find $r=0.57$, $\mu=0.618$ and
$\alpha_s=-3.5\times 10^{-2}$. Usually we cannot take so small
number of e-folds. A more reasonable number of e-folds related to
the observable perturbations is $N=50$. In order to make the
expansion for the noncommutative parameter trusted, we take
$\mu=0.5$ and we find $n_s=1.04$ and $\alpha_s=-0.002$.

\noindent 1) For n=4.

The spectral index and its running and the tensor-scalar ratio are
\eqn\cirf{\eqalign{n_s&=1+\(\mu-{3\over 16}\)r, \cr \alpha_s&=-{3r^2
\over 256}\(1+16\mu \), \cr r&={16\over N}.}} If we take $14\leq N
\leq 75$, $0.21\leq r \leq 1.14$ and $-1.5\times (1+16\mu)\times
10^{-2}\leq \alpha_s \leq -5.2\times (1+16\mu)\times 10^{-4}$. In
the commutative spacetime $\mu=0$, the running of the spectral index
satisfies $\alpha_s\in (-10^{-2}, -10^{-4})$. If we take $N=14$ to
fit $n_s=1.21$, we find $r=1.14$, $\mu=0.372$ and
$\alpha_s=-10^{-1}$. For $N=50$ and $\mu=0.5$, we find $n_s=1.1$ and
$\alpha_s=-0.01$.

The results are showed in fig 2. If we take the running spectral
index into account, the chaotic inflations with potential $V\sim
\phi^2$ and $V\sim \phi^4$ are on the boundary of the $2\sigma$
range if $\mu=0.5$ and $N=50$. If the number of the e-folds can be
released to $N=14$, these two chaotic inflation can not be ruled out
at a level of about 2 standard deviations by WMAP three years data.

\bigskip
{\vbox{{\epsfxsize=12cm
        \nobreak
    \centerline{\epsfbox{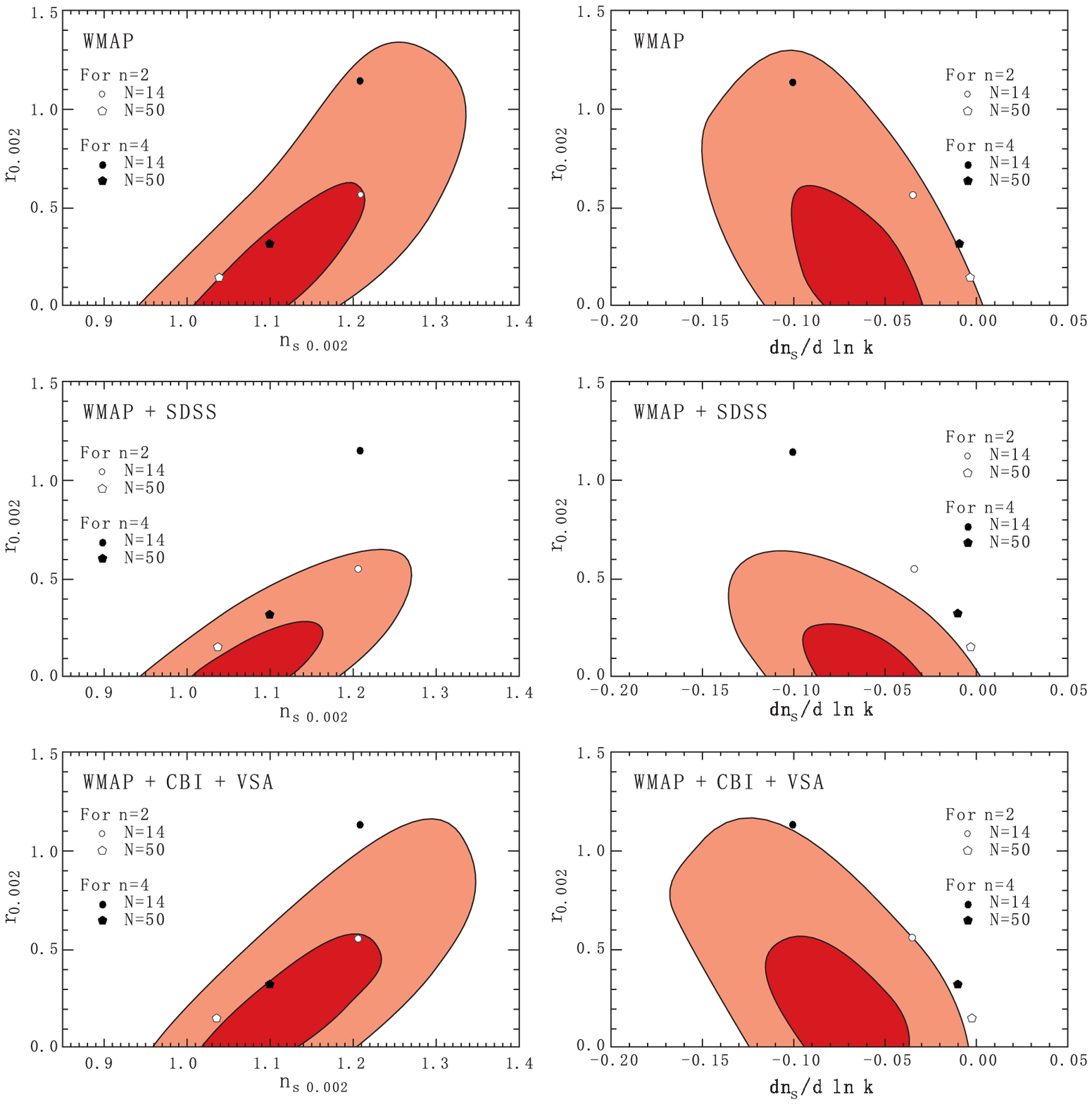}}
        \nobreak\bigskip
    {\raggedright\it \vbox{
{\bf Figure 2.} {\it With potential $V(\phi)\sim \phi^n$. $N$ is the
number of e-folds before the end of the inflation. Here we used the
copy of Fig. 12 in \wmapts. }
 }}}}
    \bigskip}

\medskip
\noindent $\bullet$ {\it Power-law inflation}

Power-law inflation is drived by a scalar field with potential
$V(\phi)=\lambda^4\exp\(-\sqrt{2\over n}{\phi \over M_p} \)$. The
scale factor can be exactly solved as \eqn\plsf{a(t)=\({t \over
(n+1)l} \)^n, } here we re-parameterize $\lambda$ to be $l$. The
amplitude of the scalar power spectrum is given by (in \refs{\hlf,
\hle}) \eqn\plmp{\Delta_{\cal R}^2=Bk^{-{2 \over
n-1}}\(1-\sigma\({k_c \over k} \)^{4 \over n-1} \), } where
\eqn\plpm{\eqalign{B&=\({n(2n-1) \over (n+1)^2} \)^{n\over n-1}{n
\over 8\pi^2}\({l_p \over l} \)^2l^{-{2 \over n-1}}, \cr
\sigma&={4n^2(n-2)(2n+1) \over (n+1)^2(n-1)(2n-1)}, \cr
k_c&=\({n(2n-1) \over (n+1)^2} \)^{n+1\over 4}l_s^{-1}\({l_s \over
l} \)^n, }} where $l_p=M_p^{-1}=8.106\times 10^{-33}$cm. The
spectral index and its running are given by \eqn\plin{n_s=1-{2
\over n-1}+{4\sigma \over n-1}\({k_c \over k} \)^{4\over n-1}, }
\eqn\plrun{\alpha_s=-{16\sigma \over (n-1)^2}\({k_c \over
k}\)^{4\over n-1} } and the tensor-scalar ratio is
\eqn\pltsr{r={16 \over n}.} Fitting the central value of the
amplitude of power spectrum, spectral index and its running in
\wm, we find $n=14.9$, $l=3.05\times 10^{-25}$ cm and
$l_s=3.85\times 10^{-29}$ cm. Or equivalently, $M_s
=l_s^{-1}=2.1\times 10^{-4} M_p$. We also find $k_c=5.38\times
10^{-5} Mpc^{-1}$ and the tensor-scalar ratio $r=1.1$ which is
within the allowed range of WMAP. Further we also calculate the
spectral index and its running at $k=0.05 Mpc^{-1}$ as $n_s=0.996$
and $\alpha_s=-0.040$.

\medskip
\noindent $\bullet$ {\it Small-field inflation}

This kind of model predicts a tiny tensor-scalar ratio. The
spacetime noncommutative effects can not help us to improve the
running spectral index significantly.

\medskip
To conclude, the three year WMAP results reveal that the CMB power
spectrum is not featureless, and a running spectral index is still
alive and the model independent analysis shows that the
noncommutative inflation model can explain the new results. However,
in some typical inflation models, such as chaotic inflation, a
rather low e-folding number is required and may appear unnatural.
The power spectrum for the power-law inflation model becomes red at
$k=0.05 Mpc^{-1}$ in our results which is consistent with WMAP. But
the running seems a little too large than what we expect. Thus it is of
great interest to have an investigation of a more general mechanism
to generate a large running. We hope to return to this problem in
the near future.

\bigskip

Acknowledgments.

The work of QGH was supported by a grant from NSFC, a grant from
China Postdoctoral Science Foundation and and a grant from K. C.
Wang Postdoctoral Foundation. The work of ML was supported by a
grant from CAS and a grant from NSFC.

\listrefs
\end